\documentclass[aps, prd, 12pt, tightenlines, notitlepage, superscriptaddress, nofootinbib, floatfix]{revtex4-1}

\usepackage{graphicx,multirow}
\usepackage{amsmath,amsfonts}
\pdfoutput=1

\usepackage{bm, xcolor, xspace}
\usepackage{hyperref,breakurl}

\definecolor{red}{rgb}{1,0,0}

\newcommand{\nn}{\nonumber}
\newcommand{\be}{\begin{eqnarray}}
\newcommand{\ee}{\end{eqnarray}}
\newcommand{\beq}{\begin{equation}}
\newcommand{\eeq}{\end{equation}}
\newcommand{\beqa}{\begin{eqnarray}}
\newcommand{\eeqa}{\end{eqnarray}}

\newcommand{\ek}{\ifmmode \epsilon_K \else $\epsilon_K$\xspace\fi}
\newcommand{\ep}{\ifmmode \epsilon^\prime/\epsilon \else $\epsilon^\prime/\epsilon$\xspace\fi}
\renewcommand{\d}{{\rm d}}

\newcommand{\Bbar}{\,\overline{\!B}{}}
\newcommand{\Dbar}{\,\overline{\!D}{}}
\newcommand{\Kbar}{\,\overline{\!K}{}}
\def\B0bar{\Bbar{}^0}
\def\D0bar{\Dbar{}^0}
\def\K0bar{\Kbar{}^0}
\def\rhobar{\bar\rho}
\def\etabar{\bar\eta}

\makeatletter
\g@addto@macro\bfseries{\boldmath}
\makeatother

\begin{document}

\title{A new look at the theory uncertainty of \ek}

\author{Zoltan Ligeti}
\affiliation{Ernest Orlando Lawrence Berkeley National Laboratory,
University of California, Berkeley, CA 94720, USA}

\author{Filippo Sala}
\affiliation{LPTHE, CNRS, UMR 7589, 4 Place Jussieu, F-75252, Paris, France}

\begin{titlepage}

\begin{abstract}
\smallskip

The observable \ek is sensitive to flavor violation at some of the highest
scales. While its experimental uncertainty is at the half percent level, the
theoretical one is in the ballpark of 15\%.  We explore the nontrivial dependence
of the theory prediction and uncertainty on various conventions, like the phase
of the kaon fields.
In particular, we show how such a rephasing allows to make the short-distance
contribution of the box diagram with two charm quarks, $\eta_{cc}$, purely real.  Our
results allow to slightly reduce the total theoretical uncertainty of \ek, while
increasing the relative impact of the imaginary part of the long distance contribution,
underlining the need to compute it reliably.  We also give updated bounds on the new physics
operators that contribute to \ek. 
\end{abstract}

\maketitle

\tableofcontents

\end{titlepage}

\section{Introduction}

The study of mixing and $CP$ violation in the $K^0$\,--\,$\K0bar$ system was
crucial for the development of the standard model (SM).  The comparison of the
measurement of the $CP$ violating parameter in $K^0$\,--\,$\K0bar$ mixing,
$\ek$, with its SM calculation provides important constraints on the CKM
matrix.  The observable \ek also probes some of the highest new physics (NP)
scales, and it gives severe constraints on explicit models of flavor. Moreover,
to distinguish between possible NP interpretations of flavor anomalies, it is
particularly important to know the level of consistency between the constraints
on the flavor sector from $K$ and $B$ decay measurements.

What are the current limiting factors of the \ek sensitivity to NP?  How can
we possibly improve them, now and in the future? 
The level to which we can answer these questions will have a major impact on our
understanding of flavor.
These limiting factors have to be looked for in the SM prediction of \ek, whose
uncertainty is more than an order of magnitude above the half percent precision
of the experimental measurement. Part of the SM uncertainty in the \ek
prediction is parametric, i.e., due to the relatively poor knowledge of some of
the CKM parameters, most notably $A$ (or equivalently~$|V_{cb}|$). This
knowledge will be substantially improved by future measurements at Belle~II and
LHCb~\cite{Aushev:2010bq,Bediaga:2012py}, which will hopefully also resolve
tensions between inclusive and exclusive determinations of
$|V_{cb}|$ and $|V_{ub}|$~\cite{Agashe:2014kda}.

Besides $|V_{cb}|$, the largest uncertainty in the SM prediction for $\ek$
originates from the calculation of $\eta_{cc}$,  the QCD correction to the box
diagram with two charm quarks.  The NNLO calculation of this
quantity~\cite{Brod:2011ty} found a large correction and a poorly behaved
perturbation series, 1, 1.38, 1.87, at leading, next-to-leading, and
next-to-next-to-leading orders, respectively, and thus quoted $\eta_{cc} = 1.87
\pm 0.76$, and $|\ek| = (1.81\pm0.28)\times 10^{-3}$.  This resulted in
different groups treating $\eta_{cc}$ differently.  For example
CKMfitter~\cite{Hocker:2001xe} uses $\eta_{cc}$ quoted in
Ref.~\cite{Brod:2011ty}, whereas UTfit~\cite{Ciuchini:2000de} uses the NLO
calculation of $\eta_{cc}$~\cite{Luca:private}.
This contributes to the visibly different \ek regions in CKMfitter
and UTfit plots.  Ref.~\cite{Buras:2013raa} instead argued that $\eta_{cc} =
1.70 \pm 0.21$ was a reasonable estimate, assuming the dominance of $\Delta m_K$
by the SM contribution, and using an estimate of the long-distance contribution
to $\Delta m_K$. Note also that the behavior of the perturbation series,
which matters for the uncertainty estimate of $\eta_{cc}$, is scheme dependent.
The perturbative QCD calculations of the $\eta_{ct} =
0.496(47)$~\cite{Brod:2010mj} and $\eta_{tt} = 0.5765(65)$~\cite{Buras:1990fn}
correction factors to the box diagrams with internal $tt$ and $ct$ quarks,
respectively, appear to be better behaved.

In this paper we show that one can eliminate $\eta_{cc}$ from the theoretical
prediction of \ek, by setting the contribution of that term to the mixing
amplitude, $M_{12}$, purely real.  While physical results are independent of
such conventions, numerically some dependence remains (similar to other scheme
dependences), because $M_{12}$ and $\Gamma_{12}$ are calculated using different
methods.  We discuss the implications of this choice on the SM uncertainty of
\ek and on the resulting constraints on NP, both at present  and in the future.

This paper is organized as follows: in Sec.~\ref{sec:review} we review some
definitions and formalism, making clear the approximations and
phase-dependences involved. In Sec.~\ref{sec:rephase} we show how to remove the
$\eta_{cc}$ contribution from \ek, and discuss the resulting modified
predictions for \ek. In Sec.~\ref{sec:implications} we comment on implications
for constraints on new physics. In Sec.~\ref{sec:conclusions} summarize our findings,
and conclude.

\newpage
\section{The state of the \ek art}

\label{sec:review}

\subsection{Definitions}

The neutral kaon mass eigenstates are linear combinations of $|K^0 \rangle =
|d\bar{s}\rangle$ and $|\K0bar \rangle = |\bar{d}s \rangle$.  The time evolution
of these states is described by the Schr\"odinger equation,
\beq
i\, \frac{d}{\d t} \bigg( \begin{matrix} K^0\\ \K0bar \end{matrix} \bigg) 
  = \left(M - i\, \frac{\Gamma}{2}\right) \bigg(
  \begin{matrix} K^0\\ \K0bar \end{matrix} \bigg) \,, 
\eeq
where the mass ($M$) and the decay ($\Gamma$) mixing matrices are $2\times2$
Hermitian matrices.  The mass eigenstates are usually labeled with their
lifetimes\footnote{The sign of $q$ is a convention, degenerate with the choice
of the phase $\theta = 0$ or $\pi$ in Eq.~(\ref{CPdef}).  Setting the
coefficients of $|K^0\rangle$ identical in $|K_L\rangle$ and $|K_S\rangle$, as done in
Eq.~(\ref{KL_KS}), sets another non-physical overall phase to~zero.}
\beq
|K_{S,L}\rangle = p |K^0\rangle \pm q |\K0bar\rangle\,,
\label{KL_KS}
\eeq
and they are the eigenvectors of $M - i\Gamma/2$. To write Eq.~(\ref{KL_KS}) we
have assumed $CPT$ symmetry, as we do in the rest of this paper. The
correspondence between the long / short lived and the heavy / light states is
\beq
K_L = K_{\rm heavy}\,, \qquad K_S = K_{\rm light}\,.
\eeq
Let us define
\beq
\Delta m = m_L - m_S > 0\,,
\eeq
and
\beq
\Delta \Gamma = \Gamma_L - \Gamma_S \simeq - \Gamma_S < 0\,.
\eeq
Throughout this paper we keep explicitly the $CP$ transformation phase
\beq\label{CPdef}
CP |K^0\rangle = e^{i\theta} |\K0bar\rangle \,, \qquad
CP |\K0bar\rangle = e^{-i\theta} |K^0\rangle \,, \qquad
\eeq
since both the $\theta=0$ and $\theta=\pi$ choices are often used in the
literature, and the cancellation of this is interesting to follow. The choice of
the phase $\theta$ is not to be confused with the phase convention for the kaon
and quark fields.

Let us define the decay amplitudes
\beq\label{AIdef}
A_f = \langle f| \mathcal{H} |K^0\rangle
  = |A_f|\, e^{i(\phi_f + \delta_f)} \,, \qquad
\bar{A_f}  = \langle f| \mathcal{H} |\K0bar\rangle
  = |A_f|\, e^{i(-\phi_f + \delta_f - \theta)} \,,
\eeq
where $\phi_f$ and $\delta_f$ are the weak and strong phases respectively, and
the amplitude ratios\footnote{The definition $\eta_f = \langle f | \mathcal{H} |
K_L\rangle / \langle f| \mathcal{H} | K_S \rangle$ is often used in the
literature, and measured magnitudes and phases are quoted.  However, there is an
arbitrary unphysical relative phase between $|K_L\rangle$ and $|K_S\rangle$. 
Effectively Eq.~(\ref{etaIdef}) is measured in the interference of
$|K_L\rangle$ and $|K_S\rangle$ decays in regeneration experiments.}
\beq\label{etaIdef}
\eta_f \equiv \frac{\langle f | \mathcal{H} | K_L\rangle}
  {\langle f| \mathcal{H} | K_S \rangle}\,
  \frac{\langle K^0|K_S\rangle}{\langle K^0|K_L \rangle}
  = \frac{1 - (q/p)(\bar A_f/A_f)}{1 + (q/p)(\bar A_f/A_f)} \,.
\eeq
In terms of $\eta_f$ for $f=\pi^+\pi^-$ and $\pi^0\pi^0$, \ek and $\epsilon'$
are defined as
\beq
\ek = \frac{2 \eta_{+-} + \eta_{00}}{3} \,, \qquad
\epsilon' = \frac{\eta_{+-} - \eta_{00}}{3}\,.
\label{epsK_def_exp}
\eeq
It is $\eta_{+-}$ and $\eta_{00}$ which are measured (and \ep is extracted from
$|\eta_{00}/\eta_{+-}|^2 \simeq 1-6\, {\rm Re}(\ep)$, valid for $|\ep| \ll 1$).

For a theoretical discussion, since $K\to\pi\pi$ decays are dominated by the
isospin $I=0$ two-pion state over  $I=2$, it is convenient to define
\beq
\eta_I = \frac{\langle (\pi\pi)_I | \mathcal{H} | K_L\rangle}
  {\langle (\pi\pi)_I | \mathcal{H} | K_S \rangle}\,
  \frac{\langle K^0|K_S\rangle}{\langle K^0|K_L \rangle}\,, \qquad
\omega \equiv \frac{\langle (\pi\pi)_{I=2} | \mathcal{H} | K_S\rangle}
  {\langle (\pi\pi)_{I=0} | \mathcal{H} | K_S \rangle}\,.
\eeq
The $CP$ violating quantities $\ek$ and $\epsilon'$ can also be defined as
\beq
\ek = \eta_0 \,, \qquad
\epsilon' = \frac{\omega}{\sqrt2}\, (\eta_2 - \eta_0) \,.
\label{epsK_def}
\eeq
The definitions in Eqs.~(\ref{epsK_def_exp}) and (\ref{epsK_def}) are
equivalent up to differences of order $|\omega \epsilon'| \sim 10^{-7}$, i.e.,
to a relative error of $10^{-4}$ for \ek, and $1/22$ for $\epsilon'$ (see
Table~\ref{tab:inputs}, and use $|\omega| = |A_2/A_0| (1+ {\cal O}(|\ek|) \simeq
1/22)$.  Neglecting isospin violation, we can further write
\begin{equation}
\eta_{+-} = \frac{\eta_0 + \eta_2 \,\omega/\sqrt2}{1+\omega/\sqrt2}\,, \qquad
\eta_{00} = \frac{\eta_0 - \sqrt{2}\,\eta_2 \,\omega}{1 - \sqrt2\,\omega}\,.
\label{etapm00_step}
\end{equation}

\subsection{\ek, phase convention independently}
\label{sec:epsK_pci}

We summarize here how to express \ek in terms of the off-diagonal elements of
the mass and width mixing matrices, $M_{12}$ and $\Gamma_{12}$ (see
Refs.~\cite{Branco:1999fs, Anikeev:2001rk} for more details). We pay
attention to write expressions that are independent of the phase conventions for
the kaon and quark fields, and we state explicitly the approximations used.

The semileptonic $CP$ asymmetry
\beq
\delta_L = \frac{\Gamma(K_L\to\pi^-\ell^+\nu) - \Gamma(K_L\to\pi^+\ell^-\bar\nu)}
  {\Gamma(K_L\to\pi^-\ell^+\nu) + \Gamma(K_L\to\pi^+\ell^-\bar\nu)}
  = (3.32 \pm 0.06) \times 10^{-3}~\mbox{\cite{Agashe:2014kda}}\,,
\label{SLasymmetry}
\eeq
measures $CP$ violation in mixing, in the limit
when $A_{\pi^+\ell^-\bar\nu} =
\bar A_{\pi^-\ell^+\nu} = 0$ and $|A_{\pi^-\ell^+\nu}| = |\bar
A_{\pi^+\ell^-\bar\nu}|$.  Note that these assumptions, valid in the SM to great
accuracy, are not precisely tested yet, as the ratio $x_+ = 
A(\K0bar \to \pi^-\ell^+\nu) / A(K^0 \to \pi^+\ell^-\bar\nu)$ is only
constrained at the $10^{-3}$ level~\cite{Agashe:2014kda}.\footnote{This is
historically called the $\Delta s = \Delta Q$ rule.  In the SM it is only
violated by higher orders in the weak interaction; when we discuss NP scenarios
below, we neglect the impact of NP on tree-level SM processes.}
In this limit, the definition in Eq.~(\ref{SLasymmetry}), and solving
the eigenvalue equations imply
\beq
\delta_L = \frac{1-|q/p|^2}{1+|q/p|^2}
  = \frac{2\,{\rm Re}(\ek)}{1+|\ek|^2}
  = \frac{2\,{\rm Im}(M_{12}^* \Gamma_{12})}{4|M_{12}|^2 + |\Gamma_{12}|^2}\,,
\label{ReEpsK_deltaL}
\eeq
where we neglected relative higher orders in $|\omega|\ep$.
The expressions for the mass and width differences that follow from the 
eigenvalue equations are
\beq
\Delta m = 2|M_{12}| \,, \qquad
\Delta \Gamma = -2|\Gamma_{12}| \,,
\label{Delta_MG}
\eeq
and are valid up to relative orders $\delta_L^2$. The relative phase
between $M_{12}$ and $\Gamma_{12}$ is $\pi + {\cal O}(\delta_L)$, since
Eq.~(\ref{ReEpsK_deltaL}) implies that its sine is small, and  the eigenvalue
equation $ 4 \,{\rm Re} (M_{12}^*\Gamma_{12}) = \Delta m \,\Delta \Gamma <0$
implies that its cosine is negative.

Equations~(\ref{ReEpsK_deltaL}) and (\ref{Delta_MG}) exhaust the information
regarding kaon mixing, and Im(\ek) is related to $CP$ violation in interference
of decay with and without mixing. Still, \ek is the observable used to constrain
$CP$ violation in $K^0$ mixing.  The reason is that \ek is measured with about 4
times smaller relative uncertainty than $\delta_L$, and the phase of \ek also
depends only on mixing parameters. Indeed, the following relation for the phase
$\phi_\epsilon$,
\beq
\phi_\epsilon
\simeq \arctan \frac{2|M_{12}|}{|\Gamma_{12}|}\,,
\label{phiEpsK}
\eeq
is valid up to relative orders $\delta_L^2$ and $|\omega^2
\ep|$, and up to
ratios of amplitudes to more than two-body final states, that do not exceed a
relative contribution of $10^{-2}$ to $\phi_\epsilon$ (see
Ref.~\cite{Lavoura:1991nu} and the updated measurements in
Ref.~\cite{Agashe:2014kda} for details). The quantity $\arctan(-2 \Delta
m/\Delta \Gamma) = 43.52^\circ$ is often referred to as ``superweak phase'', and
differs from the measured value of $\phi_{\epsilon}$ by one part in $10^4$, so
that the error of Eq.~(\ref{phiEpsK}) neither exceeds that level. Using
Eq.~(\ref{ReEpsK_deltaL}) for Re(\ek) and Eq.~(\ref{phiEpsK}) for
$\phi_\epsilon$ we obtain
\beq\label{beauty}
\ek = \frac{e^{i\phi_\epsilon} \sin\phi_\epsilon}2
  \arg \bigg(\! - \frac{M_{12}}{\Gamma_{12}} \bigg)
= e^{i\phi_\epsilon} \sin\phi_\epsilon\,
  \frac{{\rm Im}(-M_{12}/\Gamma_{12})}{2\,|M_{12}/\Gamma_{12}|}
= e^{i\phi_\epsilon} \cos\phi_\epsilon\, {\rm Im}(-M_{12}/\Gamma_{12})\,.
\eeq
Clearly, \ek only depends on $M_{12}/\Gamma_{12}$, which is physical, while the
phases of $M_{12}$ and $\Gamma_{12}$ separately are not. The neglected higher
order terms in Eq.~(\ref{beauty}) are also independent of phase conventions.
 
The standard model predictions for $M_{12}$ and $\Gamma_{12}$ are calculated
separately, using different methods, resulting in intermediate steps that depend
on phase conventions.  (In contrast, in the case of $B^0$ and $B_s^0$ mixing,
both $M_{12}$ and $\Gamma_{12}$ are computed by perturbative QCD methods, hence
the cancellations of conventions is more apparent.  In $K^0$ mixing, the use of
chiral perturbation theory, and the separate estimation of short and long
distance contributions obscure the cancellations.)  The conventions that lead to
the ``usual'' \ek formula is reviewed in the rest of this section.  In
Sec.~\ref{sec:rephase} we use the freedom of this choice to study and minimize
the uncertainties of \ek.

\subsection{\ek in the standard phase convention}
\label{sec:epsK_standard_pc}

To connect the phase convention independent manifestly physical expressions in
Eq.~(\ref{beauty}) to actual calculations, we need to consider how $M_{12}$ and
$\Gamma_{12}$ are computed. They are given by
\beq
M_{12} = \frac{1}{2 m_K}\langle K^0|\mathcal{H}|\K0bar \rangle\,, \qquad
\Gamma_{12} = \sum_f A^*(K^0 \to f) A(\K0bar \to f)\,,
\label{M12_def}
\eeq
where $f$ denote common final states of $K^0$ and $\K0bar$ decay. Usually
$M_{12}$ is written as the short-distance calculation combined with the matrix
element of the four-quark operator $O_1 = (\bar d_L\gamma_\mu s_L)^2$ in
the vacuum insertion approximation, times a ``bag parameter", $B_K$, plus
corrections.  The definition of $B_K$ involves $\theta$ via~\cite{Branco:1999fs,
Grossman:1997xn}
\beq\label{BKdef}
\langle K^0| (\bar d_L\gamma_\mu s_L) (\bar d_L\gamma^\mu s_L) |\K0bar\rangle
  = -e^{-i\theta}\, \frac 23\, B_K(\mu) f_K^2 m_K^2 \,,
\eeq
where $B_K(\mu)$ is the usual positive real quantity.  One further defines
$\widehat B_K$, to remove the $\mu$-dependence of $B_K(\mu)$.
The width mixing, $\Gamma_{12}$, is dominated by
\beq
A_0^* \bar A_0 = e^{-i\theta}\, |A_0|^2\, e^{-2i\phi_0}\,,
\label{def_Gamma12}
\eeq %
while the subleading contributions are suppressed by $|A_2/A_0|^2 \simeq 2 \times
10^{-3}$ and ${\cal B}(K_S \to f \neq \pi\pi) < 10^{-3}$.
Equations~(\ref{BKdef}) and (\ref{def_Gamma12}) show that $\theta$ drops out of
$M_{12}/\Gamma_{12}$, as it must.

In an often used $CP$ phase convention which we also use hereafter,
$\theta = \pi$~\cite{Buras:1998raa}, and
then with the usual CKM phase conventions~\cite{Agashe:2014kda}, $M_{12}$ is
near the positive real axis and $\Gamma_{12}$ is near the negative real axis. 
The weak phase, $\phi_0$, of the isospin-zero amplitude, $A_0$, depends on
hadronic matrix elements of several operators in the effective Hamiltonian.  It
is convenient and customary to define
\beq
\xi = \frac{{\rm Im}(A_0\, e^{-i\delta_0})}{{\rm Re}(A_0\, e^{-i\delta_0})}\,.
\label{xi_def}
\eeq
Without specifying phase conventions, $\xi$ can take any values between
$-\infty$ and $+\infty$, because $\phi_0$ is convention dependent. In phase
conventions in which $|\xi|\ll 1$ and ${\rm Re} (A_0\, e^{-i\delta_0}) > 0$, one
has $\xi = \arg (A_0\, e^{-i\delta_0}) = - \frac12 \arg (-\Gamma_{12})$ up to
relative orders $\xi^2$ (in addition to the phase-independent relative orders
${\cal B}(K_S \to f \neq \pi\pi)$ and $|A_2/A_0|^2$).  Then
\beq
\arg(-M_{12}/\Gamma_{12}) = \arg(M_{12}) - \arg(-\Gamma_{12})
  \simeq \frac{2\,{\rm Im}M_{12}}{2 |M_{12}|} + 2\xi \,,
\eeq
is valid to the required accuracy in phase conventions satisfying
$\{ \arg M_{12}\,,\, \arg\Gamma_{12} \} = {\cal O}(\delta_L) \ll 1$ (mod $\pi$).
Thus, starting from the manifestly convention independent Eq.~(\ref{beauty}),
choosing $\theta = \pi$ and weak phases such that $|\xi|\ll 1$, we recover the
often quoted expression,
\beq\label{usual}
\ek = e^{i\phi_\epsilon} \sin\phi_\epsilon
   \bigg( \frac{{\rm Im}M_{12}}{\Delta m} + \xi \bigg) 
= e^{i\phi_\epsilon} \sin\phi_\epsilon
   \bigg( \frac{{\rm Im}M_{12}^{\rm SD}}{\Delta m} + \xi 
   + \frac{{\rm Im}M_{12}^{\rm LD}}{\Delta m}\bigg) 
= \frac{\kappa_\epsilon\, e^{i\phi_\epsilon}}{\sqrt2} \,
  \frac{{\rm Im}M_{12}^{\rm SD}}{\Delta m} \,.
\eeq
We have explicitly separated the short-distance $\Delta s=2$ contribution,
$M_{12}^{\rm SD}$, from $\xi$, and from the long-distance contribution,
$M_{12}^{\rm LD}$.  The last term implicitly defines $\kappa_\epsilon$, which is
often written as~\cite{Buras:2010pza, Buras:2008nn}
\beq\label{kappa}
\kappa_\epsilon = \sqrt2\, \sin\phi_\epsilon
  \bigg( 1 + \rho\, \frac{\xi}{\sqrt2\, |\ek|} \bigg)\,.
\eeq

\subsection{Estimating $\xi$ and $\rho$}
\label{sec:xi_rho}

Currently available estimates of $\xi$ use either lattice QCD calculations, or
the measured value of the direct $CP$-violating quantity, $\epsilon'$, or a
combination of the two.  It must be emphasized that using  $\epsilon'$ as an
input is only valid assuming that it is determined by the SM.  (As discussed
below, it is possible that $\epsilon'$ is affected by NP but \ek is not, and
vice versa.)

One can write $\epsilon'$ as%
\beq\label{epsprime2}
\epsilon' = \frac{i}{\sqrt2}\, \bigg| \frac{A_2}{A_0} \bigg|\, 
  e^{i(\delta_2-\delta_0)}\, \sin(\phi_2-\phi_0)\,,
\eeq
valid up to relative orders $|A_2/A_0|$ and $|\ek|$. 
This expression is phase convention independent, as $\phi_2-\phi_0$ and
$\delta_2-\delta_0$ are physical, and correctly implies $\phi_{\epsilon'} =
\pi/2+\delta_2-\delta_0 = (42.3\pm1.5)^\circ$.  In phase
conventions in which $\phi_0$ and $\phi_2$ are both tiny,
\beq
\label{epsprime_pcd}
\epsilon' = \frac{e^{i\phi_{\epsilon'}}}{\sqrt2}\, \bigg| \frac{A_2}{A_0} \bigg|\, 
  \bigg[ \frac{{\rm Im}(A_2\, e^{-i\delta_2})}{|A_2|} - \xi \bigg] .
\eeq
This yields
\beq\label{xiestimate}
\xi = \frac{{\rm Im}(A_2\, e^{-i\delta_2})}{|A_2|} 
 - \sqrt2\, |\ek|\, \bigg| \frac{A_0}{A_2} \bigg|\,
 \bigg|\frac{\epsilon'}\ek \bigg| \,,
\eeq
where the relative errors in both Eqs.~(\ref{epsprime_pcd}) and
(\ref{xiestimate}), which depend on the phase convention, are of order $\xi^2$. 
The second term in Eq.~(\ref{xiestimate}) is well-known experimentally, and this
expression allows using lattice calculations of $A_2$ instead of $A_0$ to
estimate $\xi$.

Using the lattice QCD result ${\rm Im}(A_2\, e^{-i\delta_2}) = -
6.99(0.20)(0.84) \times 10^{-13}$ GeV~\cite{Blum:2015ywa}, we obtain
\beq
\xi = - (1.65 \pm 0.17) \times 10^{-4} \qquad {\rm (input~from~\ep~measurement)}.
\eeq
In contrast, the lattice calculation ${\rm Im}(A_0\,e^{-i\delta_0}) =
1.90(1.22)(1.04)\times 10^{-11}$~GeV~\cite{Bai:2015nea}, using 
Eq.~(\ref{xi_def}), yields
\beq
\xi = - (0.57 \pm 0.48) \times 10^{-4}\qquad {\rm (no~input~from~\ep~measurement)}.
\eeq
This difference is equivalent to the statement that the lattice QCD
calculations~\cite{Blum:2015ywa, Bai:2015nea} show about a $2.5\sigma$ tension
with $\epsilon'$, which can be further sharpened using additional
inputs~\cite{Buras:2015yba}.

From Eqs.~(\ref{usual}) and (\ref{kappa}), the parameter $\rho$ is defined as
\begin{equation}
\rho = 1 + \frac{1}{\xi}\, \frac{{\rm Im}(M_{12}^{\rm LD})}{\Delta m}\,.
\label{def_rho}
\end{equation}
Without a lattice computation of $M_{12}^{\rm LD}$, $\rho$ can be estimated in
the framework of chiral perturbation theory ($\chi$PT)~\cite{Buras:2010pza} (see
also~\cite{Nir:1992uv, Andriyash:2003ym}). First, one argues that all important
dispersive diagrams share the same phase~\cite{Nir:1992uv,Buras:2010pza}, so
that the phase of the absorptive and dispersive parts are related via
\begin{equation}
\frac{{\rm Im} M_{12}^{\rm LD}}{ {\rm Re} M_{12}^{\rm LD}}
  \simeq \frac{{\rm Im} \Gamma_{12}^{\rm LD}}{ {\rm Re} \Gamma_{12}^{\rm LD}} 
  \simeq - 2 \xi (1\pm 0.5)\,.
\label{M12LDxi}
\end{equation}
Here we keep using the 50\% uncertainty quoted in Ref.~\cite{Buras:2010pza} to
account for the non-aligned contributions. The dominant contribution to ${\rm
Re} M_{12}^{\rm LD}$ comes from the $\pi \pi$ loop, which has been estimated
as~\cite{Buras:2010pza}
\begin{equation}
\frac{{\rm Re} M_{12}^{\rm LD}}{\Delta m} \simeq 
  \frac{{\rm Re} M_{12}^{(\pi\pi)}}{\Delta m} \simeq 0.2 \pm 0.1\,.
\label{ReM12LD_estimate}
\end{equation}
(Preliminary lattice calculations~\cite{Bai:2014hta} hint at a smaller role for
the $2\pi$ state than the $\chi$PT estimate; refining this is important.)
Equations~(\ref{M12LDxi}) and (\ref{ReM12LD_estimate}) finally imply
\begin{equation}
\rho = 1 - 2 (0.2\pm0.14) = 0.6 \pm 0.3\,.
\label{rho_big}
\end{equation}

\subsection{Short distance contribution and usual evaluation of \ek}

Given Eqs.~(\ref{usual}) and (\ref{kappa}) and estimates of $\xi$ and $\rho$,
the only remaining ingredient in making a SM prediction for \ek is the
expression for the short-distance contribution to $M_{12}$ for $\theta = \pi$,
\begin{equation}\label{M12sd}
M_{12}^{\rm SD} = \frac{\Delta m}{\sqrt{2}}\, \widehat{C}_\epsilon \,\Big[
  \lambda_t^{*2} \eta_{tt} S_0(x_t) 
  + 2 \lambda_c^* \lambda_t^* \eta_{ct} S_0(x_t,x_c) 
  + \lambda_c^{*2} \eta_{cc} x_c \Big] ,
\end{equation}
where $\lambda_q = V_{qd} V_{qs}^*$, $x_q = [\overline m_q(\overline m_q) /
m_W]^2$, the Inami-Lim functions $S_0$ can be found, e.g., in
Ref.~\cite{Buras:1998raa}, and\footnote{The uncertainty of $\widehat C_\epsilon (= C_\epsilon \widehat B_K)$ is dominated by those of $f_K^2$ and $\widehat B_K$.  Their contributions are now comparable, making the past separation of $C_\epsilon$ and $\widehat B_K$ less motivated.
}%
\beq
\widehat{C}_\epsilon = \frac{G_F^2}{6 \sqrt2\,\pi^2}\, 
  \frac{m_K\, m_W^2}{\Delta m}\, f_K^2 \widehat{B}_K= (2.806 \pm 0.049) \times 10^4\,.
\eeq
Taking the imaginary part of $M_{12}^{\rm SD}$, we obtain from
Eq.~(\ref{usual})
\beq\label{epsold}
\ek =  \kappa_\epsilon\, e^{i\phi_\epsilon}\,\widehat{C}_\epsilon\,  |V_{cb}|^2 \lambda^2\,
  \etabar\, \Big\{ |V_{cb}|^2 \big[ (1-\rhobar) 
  + \lambda^2(\rhobar-\rhobar^2-\etabar^2) \big] \eta_{tt} S_0(x_t) 
  + \eta_{ct} S_0(x_t,x_c)
  - \eta_{cc} x_c \Big\},
\eeq
where we neglected ${\cal O}(\lambda^{14})$ terms in the CKM
expansion.\footnote{We use the expansion of the CKM matrix valid to all
orders~\cite{Hocker:2001xe}, which implies
\beqa
\lambda_c &=& - \lambda \Big[1 - \frac{\lambda^2}2 + {\cal O}(\lambda^4)\Big]
  - i\etabar A^2 \lambda^5 \Big[1 + \frac{\lambda^2}2 + {\cal O}(\lambda^4)\Big]
   \,, \nn\\*
\lambda_t &=& - A^2 \lambda^5 \Big[ 1 - \rhobar - \frac{\lambda^2}2 
  (1-3\rhobar+2\rhobar^2+2\etabar^2) + {\cal O}(\lambda^4)\Big]
  + i\etabar A^2 \lambda^5 \Big[ 1+\frac{\lambda^2}2
  + {\cal O}(\lambda^4)\Big] \,,
\eeqa}
As is usually done, we replaced $\lambda^4 A^2$ by $|V_{cb}|^2$, which is  valid
in the SM, as $V_{cb} = A\lambda^2 + {\cal O}(\lambda^8)$~\cite{Hocker:2001xe}. 
The ${\cal O}(\lambda^2)$ correction to the leading order result, proportional
to $(\rhobar-\rhobar^2-\etabar^2)$, is severely suppressed accidentally, because
$\rhobar / (\rhobar^2+\etabar^2) = \sin^2\alpha - \frac12 \sin2\alpha\cot\beta$
($\alpha$ and $\beta$ being the standard CKM angles) and $\alpha$ is near
$90^\circ$.

Below we refer to the expression for \ek in Eq.~(\ref{epsold}) as the ``usual
evaluation".  We discuss its central values and error budget together with that
of our evaluation of \ek, in Section~\ref{sec:numerics}.

\section{Removing $\eta_{cc}$ from \ek}
\label{sec:rephase}

\subsection{Rephasing the evaluation of \ek}
\label{sec:rephase_detail}

With respect to the ``standard'' phase convention that lead to
Eq.~(\ref{epsold}), one can rephase the kaon fields as
\beq
\label{rephasing_kaons}
|K^0\rangle \to |K^0\rangle' = e^{i\, {\rm arg}(\lambda_c)} |K^0\rangle, \qquad
|\K0bar\rangle \to |\K0bar\rangle' = e^{-i \,{\rm arg}(\lambda_c)} |\K0bar\rangle\,,
\eeq
which has the effect of multiplying the expression for $M_{12}^{\rm SD}$ in
Eq.~(\ref{M12sd}) by $\lambda_c^2/|\lambda_c|^2$, thus making the $\eta_{cc}$
contribution purely real.\footnote{The definition of kaons in terms of quarks
introduces two further non-physical arbitrary phases $\alpha$ and
$\tilde{\alpha}$ ($|K^0 \rangle = e^{i\alpha} |d\bar{s}\rangle$, $|\K0bar
\rangle = e^{i\tilde{\alpha}}  |\bar{d}s \rangle$). If they are set to zero,
then Eq.~(\ref{rephasing_kaons}) can also be obtained by choosing a CKM matrix
convention where $V_{cd}V_{cs}^*$ is real, e.g., $V'_{\rm CKM} = V_{\rm CKM}
\times  {\rm diag} \big(1,\, \lambda_c/|\lambda_c|,\, 1\big)$.}
Since $|{\rm Im}(\lambda_c)/{\rm Re}(\lambda_c)| < 10^{-3}$, this rephasing has a
negligible impact on the short distance contribution to $\Delta m$.
However, the impact on \ek is significant, which we study next.

All the results of Sec.~\ref{sec:epsK_pci} are still valid, being independent of
phase conventions. The results of Sec.~\ref{sec:epsK_standard_pc} and 
Eq.~(\ref{usual}) in particular are valid as well, since despite the ${\cal
O}(1)$ changes in $\arg M_{12}$ and $\arg\Gamma_{12}$, their orders of magnitude
are unchanged.  In fact, in every step the phase-dependent errors never exceed a
relative amount of ${\cal O}(\xi^2)$, and in the new phase convention $\xi' <
10^{-3}$ still holds (see below).

The consequences of the rephasing defined in Eq.~(\ref{rephasing_kaons}) are
\beqa\label{M_change}
{\rm Im} M_{12} &\to & {\rm Im} M_{12}' = {\rm Im} M_{12} 
  \frac{{\rm Re} \lambda_c^2}{|\lambda_c^2|} +
  {\rm Re} M_{12} \frac{{\rm Im} \lambda_c^2}{|\lambda_c^2|} 
  \simeq {\rm Im} M_{12} + 2\lambda^4 A^2\etabar\, {\rm Re} M_{12} \,,
\\
\label{xi_change} \xi &\to& \xi' = - \frac12 \frac{{\rm Im}(\Gamma_{12} \lambda_c^2)}
  {{\rm Re}(\Gamma_{12} \lambda_c^2)} 
\simeq - \frac{1}{2}\, \bigg( \frac{{\rm Im}\Gamma_{12}}{{\rm Re}\Gamma_{12}}
  + \frac{{\rm Im \lambda_c^2}}{{\rm Re \lambda_c^2}}\bigg)
  \simeq \xi - \lambda^4 A^2 \etabar\,.
\eeqa
Both in ${\rm Im} M_{12}'$ and in $\xi'$, the uncertainties due to neglected
terms are negligible.
Thus, the short-distance contribution to $M_{12}$ becomes
\beq\label{M12sdprime}
M_{12}^{\rm SD}{}\,' = \frac{\Delta m}{\sqrt{2}}\, \widehat{C}_\epsilon \,\bigg[
  \frac{\lambda_t^{*2}\lambda_c^2}{|\lambda_c|^2}\, \eta_{tt} S_0(x_t) 
  + 2 \lambda_c \lambda_t^*\, \eta_{ct} S_0(x_t,x_c) 
  + |\lambda_c|^2\, \eta_{cc} x_c \bigg] \,,
\eeq
and the $\eta_{cc}$ term does not contribute to the imaginary part.

For the long-distance contribution to $M_{12}$,
we can use the same estimate as in Ref.~\cite{Buras:2010pza} to obtain $\rho
= 0.6\pm 0.3$, as reviewed in Sec.~\ref{sec:xi_rho}.
We then obtain
\begin{equation}
{\rm Im} M_{12}^{\rm LD}{}\,' = -2 \big[\xi (1\pm0.5) - \lambda^4 A^2
\etabar\big]\,
  {\rm Re} M_{12}^{\rm LD} = -2(\xi' \pm 0.5\,\xi)\, {\rm Re} M_{12}^{\rm LD}\,,
\label{ImM12_prime}
\end{equation}
where in the first equality we used Eqs.~(\ref{M12LDxi}) and
(\ref{M_change}), and in the second equality Eq.~(\ref{xi_change}).
For simplicity, we define
\begin{equation}\label{kappa_prime}
\kappa_\epsilon' = \sqrt{2}\sin\phi_\epsilon \times 
  \left(1 + \rho' \frac{\xi'}{\sqrt{2} |\epsilon_K|} \right) ,
\end{equation}
with
\begin{equation}
\label{def_rho_prime}
\rho' = 1 + \frac{1}{\xi'}\frac{{\rm Im}(M_{12}^{\rm LD'})}{\Delta m}
  = 1 - 2 \left(1\pm 0.5\,\frac{\xi}{\xi'}\right) (0.2\pm0.1) \,,
\end{equation}
where in the second equality we used Eqs.~(\ref{ImM12_prime}) and
(\ref{ReM12LD_estimate}). Numerically, we find
\begin{equation}
\rho' = 0.6 \pm 0.2\,,
\end{equation}
where the uncertainty of $\rho'$ coming from the CKM inputs (contained in
$\xi'$) is negligible.

Thus, we finally obtain 
\beq\label{epsnew}
\ek =  \kappa_\epsilon'\, e^{i\phi_\epsilon} \,\widehat{C}_\epsilon \,|V_{cb}|^2 \lambda^2\,
  \etabar\, \Big\{ |V_{cb}|^2 \big[ (1-\rhobar) 
  + \lambda^2(\rhobar-\rhobar^2-\etabar^2) \big] \eta_{tt} S_0(x_t) 
  + \eta_{ct} S_0(x_t,x_c) \Big\}, 
\eeq
to which we refer below as ``our evaluation''. For convenience, we report our evaluation in a ready-to-use
form in Eqs.~(\ref{epsK_readytouse})--(\ref{kappaeps_Delta}) in
Section~\ref{sec:conclusions}.

\subsection{Numerical results and discussion}
\label{sec:numerics}

\begin{table}[tb!]
\renewcommand{\arraystretch}{1.0}
\begin{tabular}{lll}
\hline\hline
Parameter  &  value  &  source\\
\hline
$\Delta m$ &  $3.484(6) \times 10^{-12}$ MeV & \cite{Agashe:2014kda}\\
$m_{K^0}$ & $497.614(24)$ MeV & \cite{Agashe:2014kda} \\
$\Delta \Gamma$ &  $7.3382(33) \times 10^{-12}$ MeV & \cite{Agashe:2014kda}\\
$|\epsilon_K| $& $(2.228 \pm 0.011) \times 10^{-3}$ & \cite{Agashe:2014kda} \\
$\phi_\epsilon$ & $(43.52 \pm 0.05)^\circ$  & \cite{Agashe:2014kda}\\
$|\epsilon^\prime/\epsilon|$ & $(1.66 \pm 0.23) \times 10^{-3}$ & \cite{Agashe:2014kda} \\
$|A_0/A_2|$  &  $22.45(6)$& \cite{Agashe:2014kda,Liu:2012} \\
$|A_0|$ & $3.32(2) \times 10^{-7}$ GeV & \cite{Agashe:2014kda,Liu:2012} \\
\hline
$\eta_{cc}$  &  $1.87(76)$  & \cite{Brod:2011ty}\\
$\eta_{ct}$  &  $0.496(47)$  & \cite{Brod:2010mj}\\
$\eta_{tt}$  &  $0.5765(65)$  & \cite{Buras:1990fn} \\
\hline
$\overline{m}_t(\overline m_t)$  &  $162.3(2.3)$ GeV & \cite{Alekhin:2013nda}\\
$\overline{m}_c(\overline m_c)$  &  $1.275(25)$ GeV & \cite{Agashe:2014kda}\\
\hline
\raisebox{0pt}[14pt][0pt]{$\widehat{B}_K$}  & $0.7661(99)$  & \cite{Aoki:2013ldr} \\
$f_K$ & $156.3(0.9)$ MeV  & \cite{Aoki:2013ldr}\\
${\rm Im}\,(A_2 e^{-i\delta_2})$  &  $-6.99 (0.20)(0.84)\times 10^{-13}$\,GeV & \cite{Blum:2015ywa}\\
${\rm Im}\, (A_0 e^{-i\delta_0})$  &  $-1.90 (1.22)(1.04) \times 10^{-11}$\,GeV & \cite{Bai:2015nea}\\
\hline\hline
\end{tabular}
\caption{Inputs used for the calculation of \ek.}
\label{tab:inputs}
\end{table}

We collect in Table~\ref{tab:inputs} the inputs used from experimental
measurements, as well as from perturbative and lattice computations. Concerning
CKM parameters, the SM prediction of \ek is obtained using the parameters that
result from the full CKM fit. In fact, their best-fit values are practically
unaffected by the exclusion of \ek from the fit inputs~\cite{Descotes:private}.
If one wants instead to account for possible NP contributions in the CKM fit,
and obtain a prediction for \ek that is as independent as possible of such NP,
then one should use the values of the CKM parameters that come from a fit to
tree-level observables only. In this second approach, the only assumption about
NP is that it affects negligibly observables that are dominated by tree-level
processes in the SM. We show the values of the CKM parameters in these two cases
in Table~\ref{tab:CKMinputs}.\footnote{CKMfitter~\cite{Hocker:2001xe} performs
several fits, using only tree-level observables to determine $\etabar$ and
$\rhobar$.  Conservatively, we use the one where the only angle measurement
included is $\gamma(DK)$, and that combines the measured values of $|V_{ub}|$,
for consistency with our treatment of $|V_{cb}|$. CKMfitter plots the fit
results, without quoting numerical results. The values in
Table~\ref{tab:CKMinputs} are read off from the plot, which is sufficient for
our purposes, given the large uncertainties.} The increased uncertainty in
$|V_{cb}|$ and $\bar{\eta}$, when not determined from the CKM fit, reflects the
tension between exclusive and inclusive determinations of $|V_{cb}|$
and~$|V_{ub}|$.

\begin{table}[t]
\renewcommand{\arraystretch}{1.0}
\begin{tabular}{c|ccl}
\hline\hline
CKM parameters  &  SM CKM fit~\cite{Hocker:2001xe} &  tree-level only \\
\hline
$\lambda$                         &  $0.22543 \pm 0.00037$     &   $0.2253 \pm 0.0008$ & \cite{Agashe:2014kda}\\
$|V_{cb}| (= A\lambda^2)$ &  $(41.80\pm 0.51)\times 10^{-3}$  &  $(41.1\pm1.3)\times 10^{-3}$& \cite{Agashe:2014kda}\\
$\etabar$                       &  $0.3540 \pm 0.0073$                &  $0.38\pm0.04$& \cite{Hocker:2001xe} \\
$\rhobar$                       &  $0.1504\pm0.0091$            &  $0.115\pm0.065$ & \cite{Hocker:2001xe} \\
\hline\hline
\end{tabular}
\caption{The CKM parameters used as inputs.  Using the SM CKM fit results
assumes that the SM determines all observables.  The tree-level inputs are
applicable even if TeV-scale new physics affects the loop-mediated processes.}
\label{tab:CKMinputs}
\end{table}

Thus, the usual evaluation Eq.~(\ref{epsold}) and our evaluation
Eq.~(\ref{epsnew}) lead to the SM predictions for \ek shown in
Table~\ref{tab:epsKTOTAL_paper}. When interested in the SM prediction for \ek,
we use the more precise value of $\xi$, determined using the measured value of
\ep as an input (in line with the assumption that the SM accounts for all flavor
measurements). In the determination where we allow for NP, instead, we use the
lattice value of Im$(A_0)$ to determine $\xi$, instead of the measured \ep. For
convenience, we also report in Table~\ref{tab:epsKTOTAL_paper} the values of
$\xi$, $\kappa_\epsilon$ and $\xi'$, $\kappa_\epsilon'$ in our evaluation that
correspond to these choices. Finally, the various sources of uncertainties in
\ek and their relative impacts are shown in Table~\ref{tab:errorbudget_NEW}. The
total error of \ek is obtained by adding all contributions in quadrature.

\begin{table*}[tb!]
\renewcommand{\arraystretch}{1.0}
\begin{tabular}{cc|c|cc}
\hline\hline
& CKM inputs  &$|\ek| \times 10^3$ & 
  \raisebox{0pt}[14pt][0pt]{$\kappa_\epsilon^{(\prime)}$}  & 
  $\xi^{(\prime)} \times 10^4$\\
\hline
\multirow{2}{*}{Usual evaluation}
&  tree-level   &$2.30\pm 0.42$  &  $0.963 \pm 0.010$  & $-0.57 \pm 0.48$\\
&  SM CKM fit  &  $2.16 \pm 0.22$  &  $0.943 \pm 0.016$  & $-1.65 \pm 0.17$\\
\hline
\multirow{2}{*}{Our evaluation}
&  tree-level  & $2.38 \pm 0.37$  &  $0.844 \pm 0.044$  & $-6.99 \pm 0.92$\\
&  SM CKM fit  & $2.24 \pm 0.19 $  &  $0.829 \pm 0.049$  & $-7.83 \pm 0.26$\\
\hline\hline
\end{tabular}
\caption{Present value of \ek in the usual evaluation (upper part) and in our
evaluation (lower part). For convenience, we also show the values of the
quantities $\kappa_\epsilon$ and $\xi$ defined in Eqs.~(\ref{kappa}) and
(\ref{xi_def}) in the upper part, and $\kappa_\epsilon'$ and $\xi'$ defined in
Eqs.~(\ref{kappa_prime}) and (\ref{xi_change}) in the lower part.}
\label{tab:epsKTOTAL_paper}
\end{table*}

\begingroup
\squeezetable
\begin{table*}[tb!]
\renewcommand{\arraystretch}{1.0}
\tabcolsep 5.8pt
\begin{tabular}{cc|cccccc|ccc|c}
\hline\hline
&CKM inputs  &  $\eta_{cc}$ & $\eta_{ct}$  &  $\kappa_\epsilon^{(\prime)}$  &
  $m_t$  &  $m_c$  &  \raisebox{0pt}[10pt][0pt]{$f_K^2 \widehat{B}_K$}  
  &  $|V_{cb}|$ &  $\etabar$  &  $\rhobar$
  &  $|{\Delta \epsilon_K}/{\epsilon_K}|_{\rm tot.}$ \\[2pt]
\hline
\multirow{2}{*}{Usual evaluation}
&  tree-level  & 7.3\%& 4.0\%& 1.1\%& 1.7\% &  0.8\% & 1.7\%
  &11.1\%  &10.4\%  & 5.4\%  & 18.4\%\\
&  SM CKM fit   & 7.4\%  & 4.0\%  &  1.7\% & 1.7\%  &  0.8\% & 1.7\%
  & 4.2\%  &  2.0\%  &  0.8\%  & 10.2\%\\
\hline
\multirow{2}{*}{Our evaluation}
&  tree-level      &  ---  &  3.4\%  &  5.2\%  & 1.5\%& 1.2\%   & 1.7\% & 9.5\%  &  8.9\%  &  4.5\%  & 15.6\%\\
&  SM CKM fit   &  ---  & 3.4\%  &  5.9\%  &  1.5\%  &  1.3\% & 1.7\%& 3.6\%  &  1.7\%  &  0.7\%  & 8.4\%\\
\hline\hline
\end{tabular}
\caption{The present error budget of $\epsilon_K$ in the usual evaluation (upper
part) and using our evaluation (lower part).  The parameters with a
corresponding uncertainty above 1\% are shown.}
\label{tab:errorbudget_NEW}
\end{table*}
\endgroup

As expected, the central values of \ek in Table~\ref{tab:epsKTOTAL_paper} vary
according to the strategy used to compute \ek (our vs.\ usual evaluation, and
SM CKM fit vs.\ tree-level inputs).  
The central values are actually all within $1\sigma$ of each other, and of the
experimental central value $|\ek|^{\rm (exp)} = 2.228 \times 10^{-3}$.  Note
that the latest determination of $V_{cb}$ from $B\to D\ell\bar\nu$,  $|V_{cb}| =
40.8(1.0) \times 10^{-3}$~\cite{Bazavov:2016nty}, reduces the tension with its
inclusive determination (however, that from $B\to D^*\ell\bar\nu$ remains lower;
see, e.g., Ref.~\cite{Bailey:2015tba} for more discussions).
Table~\ref{tab:epsKTOTAL_paper} also shows that in our evaluation the
uncertainty in the long distance contribution to \ek (i.e.,
$\kappa_\epsilon^{(\prime)} \neq 1$) is relatively more important than in the
usual evaluation. In the latter case, the $\eta_{cc}$ term contributes to \ek
with a negative sign, and its removal in our evaluation is compensated by an
increase in the imaginary part of the long-distance contribution.
Table~\ref{tab:errorbudget_NEW} makes the usefulness of our evaluation of \ek
manifest:
\begin{itemize}
\item[$\diamond$] Given state-of-the-art inputs, our evaluation
Eq.~(\ref{epsnew}) slightly reduces the relative uncertainties of \ek with
respect to the usual one in Eq.~(\ref{epsold});

\item[$\diamond$] The gain in relative uncertainty from the removal of
$\eta_{cc}$ is partially compensated by an increase in the uncertainty from
$\kappa_\epsilon$, which is dominated by the uncertainty of the
long-distance contribution Im$(M_{12}^{\rm LD})$.
(See sections~\ref{sec:xi_rho} and~\ref{sec:rephase_detail} for its estimate,
in the usual and in our evaluation respectively.) 
\end{itemize}

These observations highlight the importance of achieving
a better theoretical control of the long-distance contribution to $M_{12}$.
While some progress could already be attained with tools like $\chi$PT,
a significant step forward probably requires an effort from lattice QCD (recent attempts in this direction have
appeared in Refs.~\cite{Christ:2012se, Bai:2014cva, Bai:2014hta}).
The importance of such an effort is even greater considering future prospects
for the \ek uncertainty, which, with the removal of $\eta_{cc}$, is dominated by
the CKM parameters. Within the next decade it should be possible to measure
$|V_{cb}|$ with an uncertainty of about $0.3 \times 10^{-3}$~\cite{Charles:2013aka, Aushev:2010bq},
to be compared with $1.3 \times 10^{-3}$ in Table~\ref{tab:CKMinputs}. This would
correspond to a reduced contribution to the \ek error budget,
\beq
\left|\frac{\Delta \epsilon_K}{\epsilon_K}\right|_{\Delta|V_{cb}| = 0.3\times 10^{-3}} = 2.2\%\,,
\label{future_epsK}
\eeq
in our evaluation of \ek ($2.6\%$ in the usual one).
Similarly, tree-dominated measurements will determine $\gamma$ and $|V_{ub}|$
with much better precision~\cite{Charles:2013aka, Aushev:2010bq,
Bediaga:2012py}, which will translate to an uncertainty of \ek due to CKM
elements comparable to the current SM CKM fit in
Table~\ref{tab:errorbudget_NEW}.
%where especially $\etabar$ has a big impact on the \ek uncertainty.
Finally, different lattice QCD calculations of $\widehat{B}_K$ obtain different
results for its uncertainty~\cite{Blum:2014tka, Carrasco:2015pra, Jang:2015sla},
which, however, do not exceed the 2--3\% percent level and are thus subdominant
in the error budget of \ek. (A more acute tension is present for the bag
parameters of non-SM operators, see Sec.~\ref{sec:implications}.)

\subsection{Further comments on the rephasing}
\label{sec:further_comments}

We collect here some remarks that are not strictly necessary to the previous
discussion, but that might help to make it clearer.
\begin{itemize}

\item[$\circ$] Looking at Table~\ref{tab:epsKTOTAL_paper}, it may appear
counterintuitive that larger $\xi$ uncertainties correspond to more precise
values of $\kappa_\epsilon$. That is the case because, when the $\xi$
uncertainty is larger, the $\xi$ central value is accidentally smaller. The
larger impact on the $\kappa_\epsilon$ uncertainty comes from $\rho$, which
multiplies $\xi$, and so its central value also impacts the error budget.

\item[$\circ$] The rephasing of kaon and quark fields is independent of the
freedom to remove the charm or up (or top) contribution, via unitarity of the
CKM matrix. The standard choice is to eliminate the $u$-quark contribution,
$\lambda_u = -\lambda_t - \lambda_c$, which we also followed. The possibility to
use CKM unitarity to remove $\lambda_c$, instead of $\lambda_u$, has been
emphasized in Ref.~\cite{Christ:2012se} (see Appendix~A of that paper). With
that choice, $M_{12}^{\rm SD}$ contains terms proportional to $\lambda_t^{*2}$,
$\lambda_u^{*2}$ and $\lambda_t^* \lambda_u^*$, and the second one will not
contribute to \ek, since $\lambda_u$ is real in the standard phase convention.

However, the expression for \ek obtained using $\lambda_c = -\lambda_t -
\lambda_u$ cannot yet be used to make precise predictions, since the
coefficients analogous to $\eta_{tt}$ and $\eta_{ct}$ have not been computed.
Ref.~\cite{Christ:2012se} argued that they would not have large uncertainties,
and that the related lattice calculations would become more accurate, due to the
suppression of the perturbative contribution for momenta smaller than $m_c$.
While this could justify pursuing that path, using $\lambda_c = -\lambda_t -
\lambda_u$ renders the top contribution sensitive to the $m_c$ scale, which is
generically associated with larger uncertainties. Our evaluation relies instead
on well established results, and allows immediate quantitative predictions.

\item[$\circ$] One may wonder if a rephasing other than that in
Eq.~(\ref{rephasing_kaons}) could reduce the \ek uncertainty even further.
Instead of Eq.~(\ref{rephasing_kaons}), an optimal choice might reduce but not
eliminate the $\eta_{cc}$ contribution to ${\rm Im}M_{12}^{\rm SD}$, and the
combined uncertainty due to $\eta_{cc}$ and $\kappa_\epsilon$ may decrease. To
explore this, let us define the general rephasing
\beq
\label{rephasing_optimal}
|K^0\rangle \to |K^0\rangle' = e^{i \,a\, {\rm arg}(\lambda_c)} |K^0\rangle, \qquad
|\K0bar\rangle \to |\K0bar\rangle' = e^{-i \,a\, {\rm arg}(\lambda_c)} |\K0bar\rangle\,,
\eeq
where the usual evaluation corresponds to $a=0$, and our evaluation to $a=1$. We
can choose $a$ to minimize the total uncertainty of \ek. We find that the
optimal values are $a \approx 1.0$ and $a \approx 0.7$ for the cases of
tree-level and SM CKM fit inputs, respectively. The resulting total
uncertainties for the latter case is $\left|\Delta
\epsilon_K/\epsilon_K\right|_{\rm total} = 7.9\%$, to be compared with 8.4\% of
the case $a=1$ in Table~\ref{tab:errorbudget_NEW}. The corresponding central \ek
value is $2.23\times 10^{-3}$.

If one considers a different evaluation of the long-distance contribution (\textit{e.g.},
assigning larger uncertainties to the estimates in Eqs.~(\ref{M12LDxi}) and/or (\ref{ReM12LD_estimate})),
then the phase conventions just discussed might not be the optimal ones.
In that case, one can find the phase redefinition that optimises the \ek error, along the lines we discussed.

\end{itemize}

\section{Constraints on new physics}
\label{sec:implications}

If a pattern of deviations from the SM is given, like in a specific model of flavor, 
then the correct strategy to study flavor and $CP$ constraints would be to perform a
fit to the SM + NP parameters (see, e.g., Ref.~\cite{Charles:2013aka}).
Here we would like to derive consequences for NP that are of a more
general validity, and do not need the specification of a model. Therefore, we
take an effective field theory (EFT) approach, and comment on explicit NP
models at the end of this section. We parametrize the NP contribution to $K^0$
mixing in terms of dimension-six operators, suppressed by a mass scale squared,
$\Lambda^2$.  The operator basis we consider consists of $O_1$, defined
before Eq.~(\ref{BKdef}), and 
\beq
O_2 = (\bar d_R s_L)^2, \quad 
O_3 = (\bar d_R^\alpha s_L^\beta)(\bar d_R^\beta s_L^\alpha)\,, \quad
O_4 = (\bar d_R s_L)(\bar d_L s_R)\,, \quad 
O_5 = (\bar d_R^\alpha s_L^\beta)(\bar d_L^\beta s_R^\alpha)\,,
\label{NP_operators}
\eeq
where $\alpha,\beta$ are color indices, that are implicit when their contraction
is between Lorentz-contracted fields.
The observable most sensitive to $O_{1,\dots,5}$ is \ek, so our
procedure is consistent ($\Delta m$, also sensitive to NP in $K^0$ mixing,
suffers from larger long-distance and $\eta_{cc}$ uncertainties).

To derive bounds on the operators in Eq.~(\ref{NP_operators}), we need both
their matrix elements between two kaon states at a certain low scale $\mu$, and
the running of their Wilson coefficients from $\Lambda$ down to that scale. The
matrix elements are defined in terms of the bag parameters, with $B_1 = B_K$ of
Eq.~(\ref{BKdef}), and
\beq
\langle K^0| O_i (\mu) |\K0bar\rangle = \frac{a_i}{4}\, B_i (\mu)\, 
  \frac{m_K^4 f_K^2}{[m_s(\mu) + m_d(\mu)]^2}\,,
\qquad i = 2, \dots, 5\,,
\label{Bag_pars}
\eeq
with $a_i = \{-5/3,\, 1/3,\, 2,\, 2/3\}$.
Recent calculations obtained partly consistent results~\cite{Boyle:2012qb,Jang:2014aea,
Carrasco:2015pra, Hudspith:2015wev}, while a 30--40\% tension between calculations
of $B_4$ and $B_5$ remains (as it was already the case nearly a decade
ago~\cite{Babich:2006bh, Nakamura:2006eq}). For definiteness, we use here the
values obtained in Ref.~\cite{Carrasco:2015pra} (in the $\overline{\rm MS}$
scheme), shown in Table~\ref{tab:NPinputs}, together with the
quark masses used.

\begin{table}[tb]
\renewcommand{\arraystretch}{1.0}
\begin{tabular}{cc|ccccc}
\hline\hline
\multicolumn{2}{c|}{Quark masses (at 3 GeV)}
  &  \multicolumn{5}{c}{Bag parameters (at 3 GeV)}\\
\hline
$\overline{m}_s$  &  $\overline{m}_d$  &
  $B_1$  &  $B_2$  &  $B_3$  &  $B_4$  &   $B_5$  \\
$86.5$ MeV &  $4.4$ MeV &  0.506 &
 0.46 &
 0.79 &
 0.78 &
 0.49 \\
\hline\hline
\end{tabular}
\caption{Inputs used for setting bounds on NP from \ek. Both the bag
parameters~\cite{Carrasco:2015pra} and the quark masses are in the
$\overline{\rm MS}$ scheme; the latter are obtained by NLO running from the
values at 2~GeV given in Ref.~\cite{Agashe:2014kda}.}
\label{tab:NPinputs}
\end{table}

We assume that only one operator deviates from the SM at the high scale
$\Lambda$, with a purely imaginary coefficient. We run it down to the scale $\mu
= 3$ GeV, at which the matrix elements are given.  Because of the large
uncertainties of the bag parameters $B_i$, we use the LO running and mixing of
the Wilson coefficients of $O_{1,\dots,5}$~\cite{Bagger:1997gg, Ciuchini:1997bw}
(see Refs.~\cite{Bertone:2012cu,Mescia:2012fg} for a consistent treatment
of the Wilson coefficients together with the bag parameters at NLO).

We then express the constraints from \ek as lower bounds on $\Lambda$, requiring
the NP contribution to the experimental measured value of \ek to be less than
twice the theoretical uncertainties in Table \ref{tab:errorbudget_NEW}, i.e.,
31\% for tree-level inputs and 16\% for SM CKM fit inputs (keeping in mind the
last point of Section \ref{sec:further_comments}). We ignore the differences
between the experimental central value of \ek and the theoretical predictions,
because it is small and depends anyway on the CKM parameters resulting from a
specific fit, and because this way the constraint on NP is independent of its
sign.

\begin{figure}[t]
\centering
\includegraphics[width=10 cm]{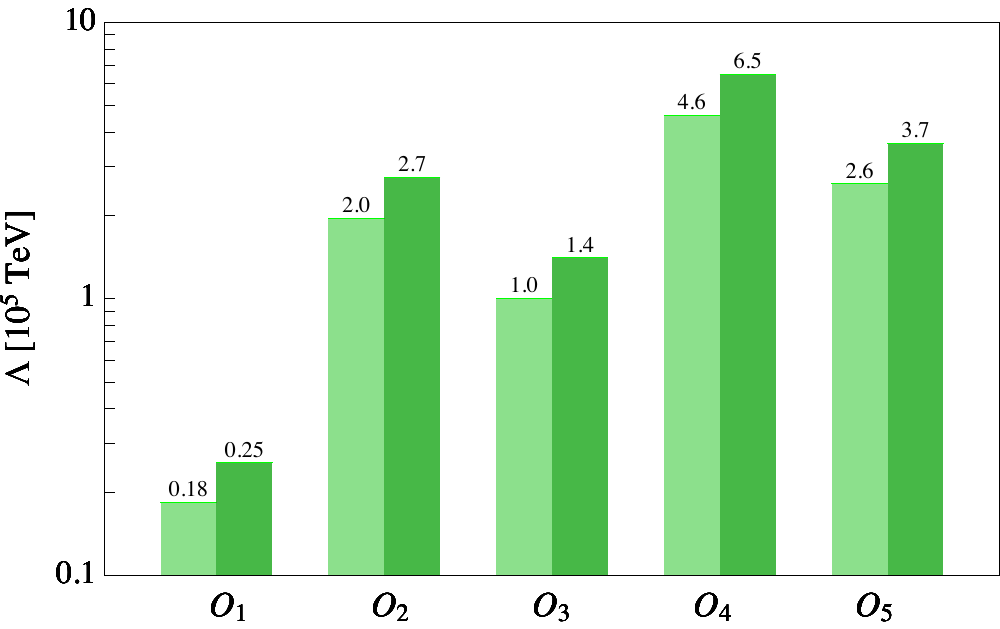}\hfill
\caption{Lower bounds from \ek on the new physics scale $\Lambda$ suppressing
each of the operators $O_{1,\dots,5}$, in units of $10^5$ TeV. For each operator
we give on the left (lighter green) the bound from tree-level CKM inputs, on the
right (darker green) the bound from SM CKM inputs.}
\label{fig:LambdaNP}
\end{figure}

The results are shown in Figure~\ref{fig:LambdaNP}, both for the SM CKM fit and
for tree-level inputs, as darker (right) and lighter (left) histograms,
respectively. From the point of view of NP, the former case assumes \ek to be
the most sensitive observable to flavor violation, and the second one is more
conservative and only requires NP not to substantially affect processes that are
tree-level in the SM. The operator most constrained by \ek is $O_4$, which
probes scales near $10^6$~TeV.

In addition we show, in Table \ref{tab:bounds_coefficients}, the resulting
bounds on the imaginary part of the Wilson coefficients $C_i$ of the operators
$\mathcal{O}_i$, for a fixed scale $\Lambda = 3$~TeV. That is useful for the
reader interested in models with new degrees of freedom not too far from the TeV
scale. In fact, the running from the scales shown in Figure~\ref{fig:LambdaNP}
down to 3 TeV is a sizable effect, which yields differences of order 50\% or
larger in the constraints on the Wilson coefficients. The same differences are,
instead, below the 10\% level if the running is performed from 3~TeV to, say, 1
or 10~TeV.

\begin{table*}[tb!]
\renewcommand{\arraystretch}{1.0}
\begin{tabular}{cr|ccccc}
\hline\hline
\multicolumn{2}{c|}{${\rm Im}(C_i)\, \dfrac{(3~{\rm TeV})^2}{\Lambda^2} < X$ $\phantom{=}$}  & $\mathcal{O}_1$ & $\mathcal{O}_2$ & $\mathcal{O}_3$ & $\mathcal{O}_4$ & $\mathcal{O}_5$
\\[4pt]
\hline
 tree-level	& $X=$	& $2.4 \times 10^{-8}$  & $3.3 \times 10^{-10}$  & $1.2 \times 10^{-9}$ & $7.5 \times 10^{-11}$ & $2.4 \times 10^{-10}$ \\
 SM CKM fit	& $X=$	& $1.2 \times 10^{-8}$  & $1.7 \times 10^{-10}$  & $6.2 \times 10^{-10}$ & $3.9 \times 10^{-11}$ & $1.2 \times 10^{-10}$ \\
\hline\hline
\end{tabular}
\caption{Upper bounds from \ek on the imaginary parts of the Wilson coefficients
of the operators $\mathcal{O}_{1,\dots,5}$, run down to 3 GeV from a scale of 3
TeV. For each operator we give the bound both from the tree-level CKM inputs and
from the SM CKM inputs.}
\label{tab:bounds_coefficients}
\end{table*}

We end this section with comments concerning the sensitivity of \ek to explicit
and widely studied NP flavor models:
\begin{itemize}
\item[$\diamond$] Composite Higgs models with partial compositeness (see,
e.g.,~\cite{Contino:2006nn}) constitute a case where \ek is the most sensitive
observable to flavor and $CP$ violation~\cite{Csaki:2008zd,Barbieri:2012tu},
unless a flavor symmetry is imposed on the strong sector~\cite{Redi:2011zi,
Barbieri:2012uh, Redi:2012uj, Barbieri:2012tu}. Then it is reasonable to
derive bounds from \ek using inputs from a CKM fit that assumes the SM, 
and corresponds to the $\sim 8\%$ theory error in Table \ref{tab:errorbudget_NEW}.
This procedure implies for example that, in the language of Ref.~\cite{Barbieri:2012tu}
and with an anarchic flavor structure in the strong sector,
\ek constrains composite fermion resonances to have masses larger than $\sim 30$ TeV.

\item[$\diamond$] Other motivated cases are models realizing a ``CKM-like''
pattern of flavor and $CP$ violation, with SM-like suppressions for the
operators present in the SM, and vanishing $O_{2,\dots,5}$. As argued in
Ref.~\cite{Barbieri:2014tja}, they consist either in
$U(3)^3$~\cite{Chivukula:1987py, Hall:1990ac, D'Ambrosio:2002ex}, or in
$U(2)^3$~\cite{Barbieri:2011ci, Barbieri:2012uh} models, all the other
symmetries being equivalent to them. In these models there is not a clear
hierarchy between observables in sensitivity to NP. The correct procedure to
analyze the impact of flavor and $CP$ violation is, therefore, to perform a fit
to the SM+NP flavor parameters, using the theoretical prediction of \ek (see
Eqs.~(\ref{epsK_readytouse})--(\ref{kappaeps_Delta}) for ready-to-use
expressions).

\item[$\diamond$] More specifically, while in general the scales probed
by \ek are higher than those probed by \ep, in CKM-like models an EFT analysis
shows~\cite{Barbieri:2012bh} that \ep is more sensitive to NP than \ek. However,
in concrete realizations it is not difficult to reverse this conclusion, for
example in supersymmetry with the first two generations heavier than the third
one~\cite{Barbieri:2012bh}.
\end{itemize}

\section{Conclusions and outlook}
\label{sec:conclusions}

Without any clear deviation from the CKM picture of flavor and $CP$ violation,
it is hard, if not impossible, to shed light on a more fundamental theory
of flavor. Among all observables, \ek probes some of the highest energies, and
puts some of the most severe constraints on explicit flavor models. It is
therefore important to improve its SM prediction, which has a much larger
uncertainty than its experimental determination.

The theory uncertainty of \ek depends on the uncertainty of CKM parameters, most
notably on that of $A$ (or equivalently $|V_{cb}|$). The largest non-parametric
uncertainty until now has been due to the perturbative QCD correction to the box
diagram with two charm quarks, $\eta_{cc}$. We showed that the dependence of \ek
on $\eta_{cc}$ can be removed via a rephasing of the kaon fields, which makes
this contribution to $M_{12}$ purely real.
In other words, in our phase convention, the contribution to \ek
from dimension-six operators always contains the top mass scale.
The resulting uncertainty of the SM
prediction of \ek is slightly reduced and, perhaps more importantly, the largest
source of non-parametric error now comes from the long distance contribution to
$M_{12}$. Thus, our formulation highlights the importance to achieve a better
theoretical control of the latter, possibly using lattice QCD. The case is
further strengthened by the precision with which the CKM inputs are expected to
be measured at Belle~II and LHCb.

In Section~\ref{sec:review}, we reviewed the derivation of the SM prediction for
\ek, explicitly exhibiting the phase convention dependences and the approximations
used. Our evaluation is presented in Section~\ref{sec:rephase}, together with
its numerical consequences for the central values and uncertainties of \ek
summarized in Table~\ref{tab:epsKTOTAL_paper}. The detailed error budget of \ek,
in our evaluation, is compared with the conventional one in
Table~\ref{tab:errorbudget_NEW}.

Finally, we provided updated constraints on new physics contributions to \ek in
Section~\ref{sec:implications}, taking full advantage of the rephasing freedom.
We also discussed how they apply to CKM-like models, and to composite Higgs
models with an anarchic flavor structure. The constraints in
Fig.~\ref{fig:LambdaNP} and Table~\ref{tab:bounds_coefficients} provide a
well-defined quantification of the \ek sensitivity to NP, and are obtained from
imposing
\beq
|\ek|^{\rm (NP)} < \begin{cases}
  0.31\, |\ek|^{\rm (exp)}
  \qquad \mbox{(tree-level inputs)}\,, \\[2pt]
  0.16\, |\ek|^{\rm (exp)}
  \qquad \mbox{(SM CKM fit inputs)}\,, 
\label{epsK_bounds}
\end{cases}
\eeq
as discussed in Sec.~\ref{sec:implications}.

Such an analysis ignores the pattern and
correlations typical of specific NP realizations. For the convenience of the
reader interested in such an analysis, that needs the CKM parameters coming from
its own SM + NP fit, we report here our ready-to-use expression for \ek without
$\eta_{cc}$,
\beq
\ek =  \kappa_\epsilon'\, e^{i\phi_\epsilon} \,\widehat{C}_\epsilon\,  |V_{cb}|^2 \lambda^2\,
  \etabar\, \Big\{ |V_{cb}|^2 \big[ (1-\rhobar) 
  + \lambda^2(\rhobar-\rhobar^2-\etabar^2) \big] \eta_{tt} S_0(x_t) 
  + \eta_{ct} S_0(x_t,x_c) \Big\},
\label{epsK_readytouse}
\eeq
where $\kappa_\epsilon'$ is given using either the measured \ep value as an 
input or using only SM lattice inputs by
\beq
\kappa_\epsilon' = 
\begin{cases}
0.834 - 0.11 \Delta \pm \left( 0.047 + 0.036 \Delta \right),
  \qquad \mbox{(\ep and lattice Im$(A_2)$ input)}\,, \\[2pt]
0.854 - 0.11 \Delta \pm \left( 0.041 + 0.035 \Delta \right),
  \qquad \mbox{(lattice Im$(A_0)$ input)}\,, 
\label{kappaeps_readytouse}
\end{cases}
\eeq
and
\beq
\Delta = \frac{\bar{\eta}}{0.35} \left( \frac{|V_{cb}|}
  {41\times 10^{-3}}\right)^2 - 1\,.
\label{kappaeps_Delta}
\eeq
Equations~(\ref{epsK_readytouse}) and (\ref{kappaeps_readytouse}), and the
inputs in Table~\ref{tab:inputs} (which imply $\widehat{C}_\epsilon = (2.806 \pm
0.049)\times 10^4$), allow making predictions for \ek for the preferred values
of CKM parameters $\lambda$, $|V_{cb}| = A \lambda^2$, $\etabar$, and $\rhobar$.

%%spacing from \section definition in revtex4-1.cls
\vspace{.8cm plus 1ex minus .2ex}
\centerline{\small\bf ACKNOWLEDGMENTS}
\vspace{.5cm}

We thank Joachim Brod, Robert Cahn, Nicolas Garron, Diego Guadagnoli, Gino
Isidori, Michael Luke, Aneesh Manohar, Michele Papucci, and Gilad Perez for
helpful conversations, and Teppei Kitahara for pointing out typos in a previous version of this paper. 
ZL was supported in part by the Office of Science, Office of High Energy
Physics, of the U.S.\ Department of Energy under contract DE-AC02-05CH11231.
FS~is supported by the European Research Council ({\sc Erc}) under the EU
Seventh Framework Programme (FP7/2007-2013)/{\sc Erc} Starting Grant (agreement
n.\ 278234 --- `{\sc NewDark}' project).

%\bibliography{Epsilon_K}
%\end{document}

\end{document}